\documentclass{ws-ijfcs}
\usepackage{enumerate}
\usepackage{url}
\usepackage{ogonek}
\usepackage{comment}

\usepackage{tikz}
\usetikzlibrary{decorations,decorations.pathreplacing,arrows,shapes,automata,backgrounds,fit,calc,circuits.logic.US}
\usetikzlibrary{arrows,shapes.gates.logic.US,shapes.gates.logic.IEC,calc}
\tikzstyle{branch}=[fill,shape=circle,minimum size=3pt,inner sep=0pt]

\newcommand{\Oh}[1]{\ensuremath{\mathcal{O}\!\left({#1}\right)}}

\begin{document}

\markboth{Bannai et al.}{Diverse Palindromic Factorization is NP-Complete}

%
\catchline{}{}{}{}{}
%

\title{Diverse Palindromic Factorization is NP-Complete}

%
\author{Hideo Bannai}
\address{Department of Informatics, Kyushu University, Japan\\
\email{bannai@inf.kyushu-u.ac.jp}}

\author{Travis Gagie}
\address{Diego Portales University, Chile\\
\email{travis.gagie@mail.udp.cl}}

\author{Shunsuke Inenaga}
\address{Department of Informatics, Kyushu University, Japan\\
\email{inenaga@inf.kyushu-u.ac.jp}}

\author{Juha K\"arkk\"ainen \and Dominik Kempa}
\address{Department of Computer Science and Helsinki Institute for
  Information Technology HIIT,\\
University of Helsinki, Finland\\
\email{juha.karkkainen@cs.helsinki.fi,  dominik.kempa@cs.helsinki.fi}}

\author{Marcin Pi\k{a}tkowski
}
\address{Faculty of Mathematics and Computer Science, Nicolaus Copernicus University, Poland\\
\email{marcin.piatkowski@mat.umk.pl}}


\author{Shiho Sugimoto}
\address{Department of Informatics, Kyushu University, Japan\\
\email{shiho.sugimoto@inf.kyushu-u.ac.jp}}

\maketitle

\begin{history}
\received{(Day Month Year)}
\accepted{(Day Month Year)}
\comby{(xxxxxxxxxx)}
\end{history}

\begin{abstract}
We prove that it is NP-complete to decide whether a given string can be factored into palindromes that are each unique in the factorization.
\end{abstract}



\section{Introduction}
\label{sec:introduction}

Given a string (or word) $S = S[1..n] = S[1]S[2]\ldots S[n]$ of $n$ symbols 
(or characters) drawn from an alphabet $\Sigma$, a {\em factorization} of $S$ partitions
$S$ into substrings (or {\em factors}) $F_1,F_2, \ldots F_t$, such that $S = F_1F_2\ldots
F_t$. Several papers have appeared recently on the subject of {\em palindromic 
factorization}; that is, factorizations where every factor is a palindrome. For example,
a palindromic factorization of the 10-symbol string $S = abaaaaabaa$ would be $aba$, $aa$, $aabaa$.

The palindromic length of a string is the minimum number of palindromic substrings into which the string can be factored.  Notice that, since a single symbol is a palindrome, the palindromic length of a string is always defined and at most the length of the string. For our example string above, $abaaaaaba$, $a$ is the palindromic factorization of minimum length.  Ravsky~\cite{Rav03} proved a tight bound on the maximum palindromic length of a binary string in terms of its length.  Frid, Puzynina, and Zamboni~\cite{FPZ13} conjectured that any infinite string in which the palindromic length of any finite substring is bounded, is ultimately periodic.  Their work led other researchers to consider how to efficiently compute a string's palindromic length and give a minimum palindromic factorization.  It is not difficult to design a quadratic-time algorithm that uses linear space, but doing better than that seems to require some string combinatorics.

Alatabbi, Iliopoulos and Rahman~\cite{AIR13} first gave a linear-time algorithm for computing a minimum factorization into maximal palindromes, if such a factorization exists.  Notice that \(a b a c a\) cannot be factored into maximal palindromes, for example, because its maximal palindromes are $a$, \(a b a\), $a$, \(a c a\) and $a$.  Fici, Gagie, K\"arkk\"ainen and Kempa~\cite{FGKK14} and I, Sugimoto, Inenaga, Bannai and Takeda~\cite{ISIBT14} independently then described essentially the same $\Oh{n \log n}$-time algorithm for computing a minimum palindromic factorization.  Shortly thereafter, Kosolobov, Rubinchik and Shur~\cite{KRS15} gave an algorithm for recognizing strings with a given palindromic length.  Their result can be used to compute the palindromic length $\ell$ of a string of length $n$ in $\Oh{n \ell \log \ell}$ time.  We also note that Gawrychowski, Merkurev, Shur and Uznanski~\cite{GMSU16} used similar techniques as Fici et al.\ and I et al., for finding approximately the longest palindrome in a stream.

We call a factorization {\em diverse} if each of the factors is
unique.  Some well-known factorizations, such as the LZ77~\cite{ZL77}
and LZ78~\cite{ZL78} parses, are diverse (except that the last factor
may have appeared before).  Fernau, Manea, Merca\c{s} and
Schmid~\cite{FMMS15} recently proved that it is NP-complete to
determine whether a given string has a diverse factorization of at
least a given size, and Schmid~\cite{Sch16} has investigated related
questions.  It seems natural to consider the problem of determining
whether a given string has a diverse factorization into palindromes.
For example, \(bgikkpps\) and \(bgikpspk\) each have exactly one such
factorization --- i.e., \((b,\ g,\ i,\ kk,\ pp,\ s)\) and \((b,\ g,\
i,\ kpspk)\), respectively --- but \(bgkpispk\) has none.  This
problem is obviously in NP and in this paper we prove that it is
NP-hard and, thus, NP-complete.
 
We also show --- proving a conjecture from the conference version of
this paper~\cite{BGIKKPPS15} --- that it is NP-complete for any fixed
$k$ to decide whether a given string can be factored into palindromes
that each appear at most $k$ times in the factorization; we call such
a factorization \emph{$k$-diverse}. Finally,
since several recent papers (e.g.,~\cite{BS14,CFGGS16,HLR16}) consider the effect of alphabet size on the difficulty of various string problems,
we show that the problems remain NP-complete even if the string is restricted to be binary.


\section{Outline}
\label{sec:outline}

In complexity theory, a Boolean circuit is formally a directed acyclic graph in which each node is either a source or one of a specified set of logic gates.  The gates are usually AND, OR and NOT, with AND and OR gates each having in-degree at least 2 and NOT gates each having in-degree 1.  A gate's predecessors and successors are called its inputs and outputs, and sources and sinks are called the circuit's inputs and outputs.  A circuit with a single output is said to be satisfiable if and only if it is possible to assign each gate a value true or false such that the output is true and all the gates' semantics are respected: e.g., each AND gate is true if and only if all its inputs are true, each OR gate is true if and only if at least one of its inputs is true, and each NOT gate is true if and only if its unique input is false.  Notice that with these semantics, a truth assignment to the circuit's inputs determines the truth values of all the gates.

The circuit satisfiability problem~\cite{Lev73} (see also, e.g.,~\cite{GJ79}) is to determine whether a given single-output Boolean circuit $C$ is satisfiable.  It was one of the first problems proven NP-complete and is often the first such problem taught in undergraduate courses.  We will show how to build, in time linear in the size of $C$, a string that has a diverse palindromic factorization if and only if $C$ is satisfiable.  It follows that diverse palindromic factorization is also NP-hard.  Our construction is similar to the Tseitin Transform~\cite{Tse68} from Boolean circuits to CNF formulas.

\begin{figure}[t]
\begin{center}
\begin{tabular}{ccc}
{\small NOT} & {\small AND} & {\small OR}\\[0.2cm]
\begin{tikzpicture}[label distance=2mm,baseline=-0.9cm]

    \node (i) at (0.75,0) {};
    \node (b1) [branch] at (1.25,0) {};
    \node (g1) [nand gate US, draw, thick] at (2,0) {};
    \node (o) at (3,0) {};
        
    \draw [thick] (i) -- (b1);
    \draw [thick] (b1) -- ($(g1.north west)!.5!(g1.input 1)$);
    \draw [thick] (b1) -- ($(g1.south west)!.5!(g1.input 2)$);
    \draw [thick] (g1.output) -- (o);
\end{tikzpicture}
&
\begin{tikzpicture}[label distance=2mm,baseline=-0.4cm]

    \node (i1) at (0.25,0.7) {};
    \node (i2) at (0.25,0.3) {};
    \node (g1) [nand gate US, draw, thick] at (1,0.5) {};
    \node (b1) [branch] at (1.75,0.5) {};
    \node (g2) [nand gate US, draw, thick] at (2.5,0.5) {};    
    \node (o) at (3.5,0.5) {};
        
    \draw [thick] (i1) -- ($(g1.north west)!.5!(g1.input 1)$);
    \draw [thick] (i2) -- ($(g1.south west)!.5!(g1.input 2)$);
    \draw [thick] (g1.output) -- (b1);
    \draw [thick] (b1) -- ($(g2.north west)!.5!(g2.input 1)$);
    \draw [thick] (b1) -- ($(g2.south west)!.5!(g2.input 2)$);
    \draw [thick] (g2.output) -- (o);
\end{tikzpicture}
&
\begin{tikzpicture}[label distance=2mm]
    \node (i1) at (0.75,1) {};
    \node (i2) at (0.75,0) {};
    \node (b1) [branch] at (1.25,1) {};
    \node (b2) [branch] at (1.25,0) {};
    \node (g1) [nand gate US, draw, thick] at (2,1) {};
    \node (g2) [nand gate US, draw, thick] at (2,0) {};
    \node (g3) [nand gate US, draw, thick] at (3.25,0.5) {};
    \node (o) at (4.25,0.5) {};
    
    \draw [thick] (i1) -- (b1);
    \draw [thick] (i2) -- (b2);
    \draw [thick] (b1) -- ($(g1.north west)!.5!(g1.input 1)$);
    \draw [thick] (b1) -- ($(g1.south west)!.5!(g1.input 2)$);
    \draw [thick] (b2) -- ($(g2.north west)!.5!(g2.input 1)$);
    \draw [thick] (b2) -- ($(g2.south west)!.5!(g2.input 2)$);
    \draw [thick] (g1.output) -- ($(g3.north west)!.5!(g3.input 1)$);
    \draw [thick] (g2.output) -- ($(g3.south west)!.5!(g3.input 2)$);
    \draw [thick] (g3.output) -- (o);
\end{tikzpicture}
\end{tabular}
\end{center}
\caption{Construction of NOT, AND and OR gates using NAND gates.}
\label{f:gates-nand-constr}
\end{figure}
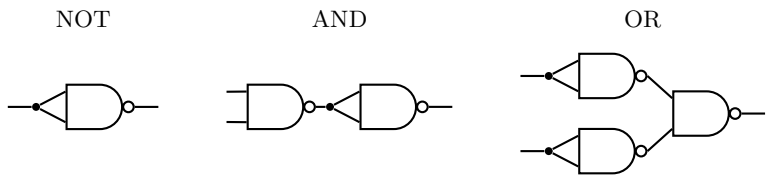

We can make each AND or OR gate's in-degree 2 and each gate's out-degree 1 at the cost of at most a logarithmic increase in the size and depth of the circuit, using splitter gates with one input and two outputs that should have the same truth value as the input.  A NAND gate is true if and only if at least one of its inputs is false.  AND, OR and NOT gates can be implemented with a constant number of NAND gates (see Fig.~\ref{f:gates-nand-constr}), so we assume without loss of generality that $C$ is composed only of NAND gates with two inputs and one output each and splitter gates.  Boolean circuits are a model for real circuits, so henceforth we assume the gates' semantics are respected, call the graph's edges wires, say each splitter divides one wire in two, and discuss wires' truth values instead of discussing the truth values of the gates at which those wires originate.

We assume each wire in $C$ is labelled with a unique symbol (considering a split to be the end of an incoming wire and the beginning of two new wires, so all three wires have different labels).  For each such symbol $a$, and some auxiliary symbols we introduce during our construction, we use as characters in our construction three related symbols: $a$ itself, $\bar{a}$ and $x_a$.  We indicate an auxiliary symbol related to $a$ by writing $a'$ or $a''$.  We write $x_a^j$ to denote $j$ copies of $x_a$.  We emphasize that, despite their visual similarity, $a$ and $\bar{a}$ are separate characters, which play complementary roles in our reduction.  We use $\$$ and $\#$ as generic separator symbols, {\em which we consider to be distinct (from each other an from all other symbols) for each use}; to prevent confusion, we add different superscripts to their different uses within the same part of the construction.

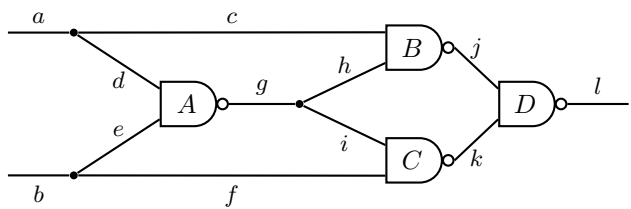
\begin{figure}[t]
\begin{center}
\begin{tikzpicture}[label distance=2mm]
    \node (i1) at (0.5,1.7) {};
    \node (i2) at (0.5,-0.2) {};
    \node (b1) [branch] at (1.5,1.7) {};
    \node (b2) [branch] at (1.5,-0.2) {};
    \node (g1) [nand gate US, draw, thick] at (3,0.75) {$A$};
    \node (b3) [branch] at (4.5,0.75) {};
    \node (g2) [nand gate US, draw, thick] at (6,1.5) {$B$};
    \node (g3) [nand gate US, draw, thick] at (6,0) {$C$};
    \node (g4) [nand gate US, draw, thick] at (7.5,0.75) {$D$};
    \node (o) at (9,0.75) {};
        
    \draw [thick] (i1) -- (b1) node [above,midway] {\small $a$}; 
    \draw [thick] (i2) -- (b2)node [below,midway] {\small $b$};
    \draw [thick] (b1) -- ($(g2.north west)!.5!(g2.input 1)$) node [above,midway] {\small $c$};
    \draw [thick] (b1) -- ($(g1.north west)!.5!(g1.input 1)$) node [below,midway] {\small $d$};
    \draw [thick] (b2) -- ($(g3.south west)!.5!(g3.input 2)$) node [below,midway] {\small $f$};
    \draw [thick] (b2) -- ($(g1.south west)!.5!(g1.input 2)$) node [above,midway] {\small $e$};
    \draw [thick] (g1.output) -- (b3) node [above,midway] {\small $g$};
    \draw [thick] (b3) -- ($(g2.south west)!.5!(g2.input 2)$) node [above,midway] {\small $h$};
    \draw [thick] (b3) -- ($(g3.north west)!.5!(g3.input 1)$) node [below,midway] {\small $i$};
    \draw [thick] (g2.output) -- ($(g4.north west)!.5!(g4.input 1)$) node [above,midway] {\small $j$};
    \draw [thick] (g3.output) -- ($(g4.south west)!.5!(g4.input 2)$) node [below,midway] {\small $k$};
    \draw [thick] (g4.output) -- (o) node [above,midway] {\small $l$};
\end{tikzpicture}
\end{center}
\caption{To construct the circuit above (computing XOR) we need to add wires $a$ and $b$, split $a$ into $c$ and $d$, split $b$ into $e$ and $f$, add gate $A$, split $g$ into $h$ and $i$, and finally add gates $B$, $C$ and $D$.}
\label{f:ex-circuit}
\end{figure}

We can build a sequence \(C_0, \ldots, C_t\) of subcircuits such that $C_0$ is empty, \(C_t = C\) and, for \(1 \leq i \leq t\), we obtain $C_i$ from $C_{i - 1}$ by one of the following operations (see Fig.~\ref{f:ex-circuit} for an example):
\begin{itemize}
\item adding a new wire (which is both an input and an output in $C_i$),
\item splitting an output of $C_{i - 1}$ into two outputs,
\item making two outputs of $C_{i - 1}$ the inputs of a new NAND gate.
\end{itemize}

We will show how to build in time linear in the size of $C$, inductively and in turn, a sequence of strings \(S_1, \ldots, S_t\) such that $S_i$ represents $C_i$ according to the following definitions:

\begin{definition}
\label{def:encoding}
A diverse palindromic factorization $P$ of a string $S_i$ {\em encodes} an assignment $\tau$ to the inputs of a circuit $C_i$ if the following conditions hold:
\begin{itemize}
\item if $\tau$ makes an output of $C_i$ labelled $a$ true, then $a$, $x_a$ and \(x_a \bar{a} x_a\) are complete factors in $P$ but $\bar{a}$, \(x_a a x_a\) and $x_a^j$ are not for \(j > 1\);
\item if $\tau$ makes an output of $C_i$ labelled $a$ false, then $\bar{a}$, $x_a$ and \(x_a a x_a\) are complete factors in $P$ but $a$, \(x_a \bar{a} x_a\) and $x_a^j$ are not for \(j > 1\);
\item if $a$ is a label in $C$ but not in $C_i$, then none of $a$, $\bar{a}$, \(x_a a x_a\), \(x_a \bar{a} x_a\) and $x_a^j$ for \(j \geq 1\) are complete factors in $P$.
\end{itemize}
\end{definition}

We say ``complete factor'' to emphasize the difference between factors in the factorization and their proper substrings; unfortunately, ``factor'' is sometimes used in the literature as a synonym for ``substring''.

\begin{definition}
\label{def:representing}
A string $S_i$ {\em represents} a circuit $C_i$ if each assignment to the inputs of $C_i$ is encoded by some diverse palindromic factorization of $S_i$, and each diverse palindromic factorization of $S_i$ encodes some assignment to the inputs of $C_i$.
\end{definition}

Once we have $S_t$, we can easily build in constant time a string $S$ that has a diverse palindromic factorization if and only if $C$ is satisfiable.  To do this, we append \(\$ \#\,x_a a x_a\) to $S_t$, where $\$$ and $\#$ are symbols not occurring in $S_t$ and $a$ is the label on $C$'s output.  Since $\$$ and $\#$ do not occur in $S_t$ and occur as a pair of consecutive characters in $S$, they must each be complete factors in any palindromic factorization of $S$.  It follows that there is a diverse palindromic factorization of $S$ if and only if there is a diverse palindromic factorization of $S_t$ in which \(x_a a x_a\) is not a factor, which is the case if and only if there is an assignment to the inputs of $C$ that makes its output true.


\section{Adding a Wire}
\label{sec:adding}

Suppose $C_i$ is obtained from $C_{i - 1}$ by adding a new wire labelled $a$.  If \(i = 1\) then we set \(S_i = x_a a x_a \bar{a} x_a\), whose two diverse palindromic factorizations \((x_a,\ a,\ x_a \bar{a} x_a)\) and \((x_a a x_a,\ \bar{a},\ x_a)\) encode the assignments true and false to the wire labelled $a$, which is both the input and output in $C_i$. If \(i > 1\) then we set
\[S_i = S_{i - 1}\,\$ \#\,x_a a x_a \bar{a} x_a\,,\]
where $\$$ and $\#$ are symbols not occurring in $S_{i - 1}$ and not equal to $a'$, $\overline{a'}$ or $x_{a'}$ for any label $a'$ in $C$.

Since $\$$ and $\#$ do not occur in $S_{i - 1}$ and occur as a pair of consecutive characters in $S_i$, they must each be complete factors in any palindromic factorization of $S_i$.  Therefore, any diverse palindromic factorization of $S_i$ is the concatenation of a diverse palindromic factorization of $S_{i - 1}$ and either \((\$,\ \#,\ x_a,\ a,\ x_a \bar{a} x_a)\) or \((\$,\ \#,\ x_a a x_a,\ \bar{a},\ x_a)\).  Conversely, any diverse palindromic factorization of $S_{i - 1}$ can be extended to a diverse palindromic factorization of $S_i$ by appending either \((\$,\ \#,\ x_a,\ a,\ x_a \bar{a} x_a)\) or \((\$,\ \#,\ x_a a x_a,\ \bar{a},\ x_a)\).

Assume $S_{i - 1}$ represents $C_{i - 1}$.  Let $\tau$ be an assignment to the inputs of $C_i$ and let $P$ be a diverse palindromic factorization of $S_{i - 1}$ encoding $\tau$ restricted to the inputs of $C_{i - 1}$.  If $\tau$ makes the input (and output) of $C_i$ labelled $a$ true, then $P$ concatenated with \((\$,\ \#,\ x_a,\ a,\ x_a \bar{a} x_a)\) is a diverse palindromic factorization of $S_i$ that encodes $\tau$.  If $\tau$ makes that input false, then $P$ concatenated with \((\$,\ \#,\ x_a a x_a,\ \bar{a},\ x_a)\) is a diverse palindromic factorization of $S_i$ that encodes $\tau$.  Therefore, each assignment to the inputs of $C_i$ is encoded by some diverse palindromic factorization of $S_i$.

Now let $P$ be a diverse palindromic factorization of $S_i$ and let $\tau$ be the assignment to the inputs of $C_{i - 1}$ that is encoded by a prefix of $P$.  If $P$ ends with \((\$,\ \#,\ x_a,\ a,\ x_a \bar{a} x_a)\) then $P$ encodes the assignment to the inputs of $C_i$ that makes the input labelled $a$ true and makes the other inputs true or false according to $\tau$.  If $P$ ends with \((\$,\ \#,\ x_a a x_a,\ \bar{a},\ x_a)\) then $P$ encodes the assignment to the inputs of $C_i$ that makes the input labelled $a$ false and makes the other inputs true or false according to $\tau$.  Therefore, each diverse palindromic factorization of $S_i$ encodes some assignment to the inputs of $C_i$.

\begin{lemma}
\label{lem:adding}
We can build a string $S_1$ that represents $C_1$.  If we have a string $S_{i - 1}$ that represents $C_{i - 1}$ and $C_i$ is obtained from $C_{i - 1}$ by adding a new wire, then in constant time we can append symbols to $S_{i - 1}$ to obtain a string $S_i$ that represents~$C_i$.
\end{lemma}


\section{Splitting a Wire}
\label{sec:splitting}

Now suppose $C_i$ is obtained from $C_{i - 1}$ by splitting an output of $C_{i - 1}$ labelled $a$ into two outputs labelled $b$ and $c$.  We set
\[S_i' = S_{i - 1}\,\$ \#\,x_a^3 b' x_a a x_a c' x_a^5\,\$' \#'\,x_a^7 \overline{b'} x_a \bar{a} x_a \overline{c'} x_a^9\,,\]
where $\$$, $\$'$, $\#$, $\#'$, $b'$, $\overline{b'}$, $c'$ and $\overline{c'}$ are symbols not occurring in $S_{i - 1}$ and not equal to $a'$, $\overline{a'}$ or $x_{a'}$ for any label $a'$ in $C$.

Since $\$$, $\$'$, $\#$ and $\#'$ do not occur in $S_{i - 1}$ and occur as pairs of consecutive characters in $S_i'$, they must each be complete factors in any palindromic factorization of $S_i'$.  Therefore, a simple case analysis shows that any diverse palindromic factorization of $S_i'$ is the concatenation of a diverse palindromic factorization of $S_{i - 1}$ and one of
\begin{eqnarray*}
&& (\$,\ \#,\ x_a^3,\ b',\ x_a a x_a,\ c',\ x_a^5,\ \$',\ \#',\ x_a^2,\ x_a^4,\ x_a \overline{b'} x_a,\ \bar{a},\ x_a \overline{c'} x_a,\ x_a^8)\,,\\
&& (\$,\ \#,\ x_a^3,\ b',\ x_a a x_a,\ c',\ x_a^5,\ \$',\ \#',\ x_a^4,\ x_a^2,\ x_a \overline{b'} x_a,\ \bar{a},\ x_a \overline{c'} x_a,\ x_a^8)\,,\\
&& (\$,\ \#,\ x_a^3,\ b',\ x_a a x_a,\ c',\ x_a^5,\ \$',\ \#',\ x_a^6,\ x_a \overline{b'} x_a,\ \bar{a},\ x_a \overline{c'} x_a,\ x_a^8)\,,\\
&& (\$,\ \#,\ x_a^2,\ x_a b' x_a,\ a,\ x_a c' x_a,\ x_a^4,\ \$',\ \#',\ x_a^7,\ \overline{b'},\ x_a \bar{a} x_a,\ \overline{c'},\ x_a^3,\ x_a^6)\,,\\
&& (\$,\ \#,\ x_a^2,\ x_a b' x_a,\ a,\ x_a c' x_a,\ x_a^4,\ \$',\ \#',\ x_a^7,\ \overline{b'},\ x_a \bar{a} x_a,\ \overline{c'},\ x_a^6,\ x_a^3)\,,\\
&& (\$,\ \#,\ x_a^2,\ x_a b' x_a,\ a,\ x_a c' x_a,\ x_a^4,\ \$',\ \#',\ x_a^7,\ \overline{b'},\ x_a \bar{a} x_a,\ \overline{c'},\ x_a^9)\,.
\end{eqnarray*}
In any diverse palindromic factorization of $S_i'$, therefore, either $b'$ and $c'$ are complete factors but $\overline{b'}$ and $\overline{c'}$ are not, or vice versa.

Conversely, any diverse palindromic factorization of $S_{i - 1}$ in which $a$, $x_a$ and \(x_a \bar{a} x_a\) are complete factors but $\bar{a}$, \(x_a a x_a\) and $x_a^j$ are not for \(j > 1\), can be extended to a diverse palindromic factorization of $S_i'$ by appending either of
\begin{eqnarray*}
&& (\$,\ \#,\ x_a^3,\ b',\ x_a a x_a,\ c',\ x_a^5,\ \$',\ \#',\ x_a^2,\ x_a^4,\ x_a \overline{b'} x_a,\ \bar{a},\ x_a \overline{c'} x_a,\ x_a^8)\,,\\
&& (\$,\ \#,\ x_a^3,\ b',\ x_a a x_a,\ c',\ x_a^5,\ \$',\ \#',\ x_a^6,\ x_a \overline{b'} x_a,\ \bar{a},\ x_a \overline{c'} x_a,\ x_a^8)\,;
\end{eqnarray*}
any diverse palindromic factorization of $S_{i - 1}$ in which $\bar{a}$, $x_a$ and \(x_a a x_a\) are complete factors but $a$, \(x_a \bar{a} x_a\) and $x_a^j$ are not for \(j > 1\), can be extended to a diverse palindromic factorization of $S_i'$ by appending either of
\begin{eqnarray*}
&& (\$,\ \#,\ x_a^2,\ x_a b' x_a,\ a,\ x_a c' x_a,\ x_a^4,\ \$',\ \#',\ x_a^7,\ \overline{b'},\ x_a \bar{a} x_a,\ \overline{c'},\ x_a^3,\ x_a^6)\,,\\
&& (\$,\ \#,\ x_a^2,\ x_a b' x_a,\ a,\ x_a c' x_a,\ x_a^4,\ \$',\ \#',\ x_a^7,\ \overline{b'},\ x_a \bar{a} x_a,\ \overline{c'},\ x_a^9)\,.
\end{eqnarray*}

We set
\[S_i = S_i'\,\$'' \#''\,x_b b x_b b' x_b \overline{b'} x_b \bar{b} x_b\,\$''' \#'''\,x_c c x_c c' x_c \overline{c'} x_c \bar{c} x_c\,,\]
where $\$''$, $\$'''$, $\#''$ and $\#'''$ are symbols not occurring in $S_i'$ and not equal to $a'$, $\overline{a'}$ or $x_{a'}$ for any label $a'$ in $C$.  Since $\$''$, $\$'''$, $\#''$ and $\#'''$ do not occur in $S_i'$ and occur as pairs of consecutive characters in $S_i'$, they must each be complete factors in any palindromic factorization of $S_i$.  Therefore, any diverse palindromic factorization of $S_i$ is the concatenation of a diverse palindromic factorization of $S_i'$ and one of
\begin{eqnarray*}
&& (\$'',\ \#'',\ x_b,\ b,\ x_b b' x_b,\ \overline{b'},\ x_b \bar{b} x_b,\ \$''',\ \#''',\ x_c,\ c,\ x_c c' x_c,\ \overline{c'},\ x_c \bar{c} x_c)\,,\\
&& (\$'',\ \#'',\ x_b b x_b,\ b',\ x_b \overline{b'} x_b,\ \bar{b},\ x_b,\ \$''',\ \#''',\ x_c c x_c,\ c',\ x_c \overline{c'} x_c,\ \bar{c},\ x_c)\,. 
\end{eqnarray*}

Conversely, any diverse palindromic factorization of $S_i'$ in which $b'$ and $c'$ are complete factors but $\overline{b'}$ and $\overline{c'}$ are not, can be extended to a diverse palindromic factorization of $S_i$ by appending
\[(\$'',\ \#'',\ x_b,\ b,\ x_b b' x_b,\ \overline{b'},\ x_b \bar{b} x_b,\ \$''',\ \#''',\ x_c,\ c,\ x_c c' x_c,\ \overline{c'},\ x_c \bar{c} x_c)\,;\]
any diverse palindromic factorization of $S_i'$ in which $\overline{b'}$ and $\overline{c'}$ are complete factors but $b'$ and $c'$ are not, can be extended to a diverse palindromic factorization of $S_i$ by appending
\[(\$'',\ \#'',\ x_b b x_b,\ b',\ x_b \overline{b'} x_b,\ \bar{b},\ x_b,\ \$''',\ \#''',\ x_c c x_c,\ c',\ x_c \overline{c'} x_c,\ \bar{c},\ x_c)\,.\]

Assume $S_{i - 1}$ represents $C_{i - 1}$.  Let $\tau$ be an assignment to the inputs of $C_{i - 1}$ and let $P$ be a diverse palindromic factorization of $S_{i - 1}$ encoding $\tau$.  If $\tau$ makes the output of $C_{i - 1}$ labelled $a$ true, then $P$ concatenated with, e.g., 
\begin{eqnarray*}
&& (\$,\ \#,\ x_a^3,\ b',\ x_a a x_a,\ c',\ x_a^5,\ \$',\ \#',\ x_a^2,\ x_a^4,\ x_a \overline{b'} x_a,\ \bar{a},\ x_a \overline{c'} x_a,\ x_a^8,\\
&& \$'',\ \#'',\ x_b,\ b,\ x_b b' x_b,\ \overline{b'},\ x_b \bar{b} x_b,\ \$''',\ \#''',\ x_c,\ c,\ x_c c' x_c,\ \overline{c'},\ x_c \bar{c} x_c)
\end{eqnarray*}
is a diverse palindromic factorization of $S_i$.  Notice $b$, $c$, $x_b$, $x_c$, \(x_b \bar{b} x_b\) and \(x_c \bar{c} x_c\) are complete factors but $\bar{b}$, $\bar{c}$, \(x_b b x_b\), \(x_c c x_c\), $x_b^j$ and $x_c^j$ for \(j > 1\) are not.  Therefore, this concatenation encodes the assignment to the inputs of $C_i$ that makes them true or false according to $\tau$.

If $\tau$ makes the output of $C_{i - 1}$ labelled $a$ false, then $P$ concatenated with, e.g.,
\begin{eqnarray*}
&& (\$,\ \#,\ x_a^2,\ x_a b' x_a,\ a,\ x_a c' x_a,\ x_a^4,\ \$',\ \#',\ x_a^7,\ \overline{b'},\ x_a \bar{a} x_a,\ \overline{c'},\ x_a^3,\ x_a^6,\\
&& \$'',\ \#'',\ x_b b x_b,\ b',\ x_b \overline{b'} x_b,\ \bar{b},\ x_b,\ \$''',\ \#''',\ x_c c x_c,\ c',\ x_c \overline{c'} x_c,\ \bar{c},\ x_c)
\end{eqnarray*}
is a diverse palindromic factorization of $S_i$.  Notice $\bar{b}$, $\bar{c}$, $x_b$, $x_c$, \(x_b b x_b\) and \(x_c c x_c\) are complete factors but $b$, $c$, \(x_b \bar{b} x_b\), \(x_c \bar{c} x_c\), $x_b^j$ and $x_c^j$ for \(j > 1\) are not.  Therefore, this concatenation encodes the assignment to the inputs of $C_i$ that makes them true or false according to $\tau$.  Since $C_{i - 1}$ and $C_i$ have the same inputs, each assignment to the inputs of $C_i$ is encoded by some diverse palindromic factorization of $S_i$.

Now let $P$ be a diverse palindromic factorization of $S_i$ and let $\tau$ be the assignment to the inputs of $C_{i - 1}$ that is encoded by a prefix of $P$.  If $P$ ends with any of
\begin{eqnarray*}
&& (\$,\ \#,\ x_a^3,\ b',\ x_a a x_a,\ c',\ x_a^5,\ \$',\ \#',\ x_a^2,\ x_a^4,\ x_a \overline{b'} x_a,\ \bar{a},\ x_a \overline{c'} x_a,\ x_a^8)\,,\\
&& (\$,\ \#,\ x_a^3,\ b',\ x_a a x_a,\ c',\ x_a^5,\ \$',\ \#',\ x_a^4,\ x_a^2,\ x_a \overline{b'} x_a,\ \bar{a},\ x_a \overline{c'} x_a,\ x_a^8)\,,\\
&& (\$,\ \#,\ x_a^3,\ b',\ x_a a x_a,\ c',\ x_a^5,\ \$',\ \#',\ x_a^6,\ x_a \overline{b'} x_a,\ \bar{a},\ x_a \overline{c'} x_a,\ x_a^8)
\end{eqnarray*}
followed by
\[(\$'',\ \#'',\ x_b,\ b,\ x_b b' x_b,\ \overline{b'},\ x_b \bar{b} x_b,\ \$''',\ \#''',\ x_c,\ c,\ x_c c' x_c,\ \overline{c'},\ x_c \bar{c} x_c)\,,\]
then $a$ must be a complete factor in the prefix of $P$ encoding $\tau$, so $\tau$ must make the output of $C_{i - 1}$ labelled $a$ true.  Since $b$, $c$, $x_b$, $x_c$, \(x_b \bar{b} x_b\) and \(x_c \bar{c} x_c\) are complete factors in $P$ but $\bar{b}$, $\bar{c}$, \(x_b b x_b\), \(x_c c x_c\), $x_b^j$ and $x_c^j$ for \(j > 1\) are not, $P$ encodes the assignment to the inputs of $C_i$ that makes them true or false according to $\tau$.

If $P$ ends with any of
\begin{eqnarray*}
&& (\$,\ \#,\ x_a^2,\ x_a b' x_a,\ a,\ x_a c' x_a,\ x_a^4,\ \$',\ \#',\ x_a^7,\ \overline{b'},\ x_a \bar{a} x_a,\ \overline{c'},\ x_a^3,\ x_a^6)\,,\\
&& (\$,\ \#,\ x_a^2,\ x_a b' x_a,\ a,\ x_a c' x_a,\ x_a^4,\ \$',\ \#',\ x_a^7,\ \overline{b'},\ x_a \bar{a} x_a,\ \overline{c'},\ x_a^6,\ x_a^3)\,,\\
&& (\$,\ \#,\ x_a^2,\ x_a b' x_a,\ a,\ x_a c' x_a,\ x_a^4,\ \$',\ \#',\ x_a^7,\ \overline{b'},\ x_a \bar{a} x_a,\ \overline{c'},\ x_a^9)
\end{eqnarray*}
followed by
\[(\$'',\ \#'',\ x_b b x_b,\ b',\ x_b \overline{b'} x_b,\ \bar{b},\ x_b,\ \$''',\ \#''',\ x_c c x_c,\ c',\ x_c \overline{c'} x_c,\ \bar{c},\ x_c)\,,\]
then $\bar{a}$ must be a complete factor in the prefix of $P$ encoding $\tau$, so $\tau$ must make the output of $C_{i - 1}$ labelled $a$ false.  Since $\bar{b}$, $\bar{c}$, $x_b$, $x_c$, \(x_b b x_b\) and \(x_c c x_c\) are complete factors but $b$, $c$, \(x_b \bar{b} x_b\), \(x_c \bar{c} x_c\), $x_b^j$ and $x_c^j$ for \(j > 1\) are not, $P$ encodes the assignment to the inputs of $C_i$ that makes them true or false according to $\tau$.

Since these are all the possibilities for how $P$ can end, each diverse palindromic factorization of $S_i$ encodes some assignment to the inputs of $C_i$.  This gives us the following lemma:

\begin{lemma}
\label{lem:splitting}
If we have a string $S_{i - 1}$ that represents $C_{i - 1}$ and $C_i$ is obtained from $C_{i - 1}$ by splitting an output of $C_{i - 1}$ into two outputs, then in constant time we can append symbols to $S_{i - 1}$ to obtain a string $S_i$ that represents $C_i$.
\end{lemma}


\section{Adding a NAND Gate}
\label{sec:nanding}

Finally, suppose $C_i$ is obtained from $C_{i - 1}$ by making two outputs of $C_{i - 1}$ labelled $a$ and $b$ the inputs of a new NAND gate whose output is labelled $c$.  Let $C_{i - 1}'$ be the circuit obtained from $C_{i - 1}$ by splitting the output of $C_{i - 1}$ labelled $a$ into two outputs labelled $a_1$ and $a_2$, where $a_1$ and $a_2$ are symbols we use only here.  Assuming $S_{i - 1}$ represents $C_{i - 1}$, we can use Lemma~\ref{lem:splitting} to build in constant time a string $S_{i - 1}'$ representing $C_{i - 1}'$.  We set
\begin{eqnarray*}
S_i'
& = & S_{i - 1}'\,\$ \#\,x_{c'}^3 a_1' x_{c'} a_1 x_{c'} \overline{a_1} x_{c'} \overline{a_1'} x_{c'}^5\\
&& \$' \#'\,x_{c'}^7 a_2' x_{c'} a_2 x_{c'} \overline{a_2} x_{c'} \overline{a_2'} x_{c'}^9\\
&& \$'' \#''\,x_{c'}^{11} b' x_{c'} b x_{c'} \bar{b} x_{c'} \overline{b'} x_{c'}^{13}\,,
\end{eqnarray*}
where all of the symbols in the suffix after $S_{i - 1}'$ are ones we use only here.

Since $\$$, $\$'$, $\$''$, $\$'''$, $\#$ and $\#'$ do not occur in $S_{i - 1}$ and occur as pairs of consecutive characters in $S_i'$, they must each be complete factors in any palindromic factorization of $S_i'$.  Therefore, any diverse palindromic factorization of $S_i'$ consists of
\begin{enumerate}
\item a diverse palindromic factorization of $S_{i - 1}'$,
\item \((\$,\ \#)\),
\item a diverse palindromic factorization of \(x_{c'}^3 a_1' x_{c'} a_1 x_{c'} \overline{a_1} x_{c'} \overline{a_1'} x_{c'}^5\),
\item \((\$',\ \#')\),
\item a diverse palindromic factorization of \(x_{c'}^7 a_2' x_{c'} a_2 x_{c'} \overline{a_2} x_{c'} \overline{a_2'} x_{c'}^9\),
\item \((\$'',\ \#'')\),
\item a diverse palindromic factorization of \(x_{c'}^{11} b' x_{c'} b x_{c'} \bar{b} x_{c'} \overline{b'} x_{c'}^{13}\).
\end{enumerate}

If $a_1$ is a complete factor in the factorization of $S_{i - 1}'$, then the diverse palindromic factorization of
\[x_{c'}^3 a_1' x_{c'} a_1 x_{c'} \overline{a_1} x_{c'} \overline{a_1'} x_{c'}^5\]
must include either
\[(a_1',\ x_{c'} a_1 x_{c'},\ \overline{a_1},\ x_{c'} \overline{a_1'} x_{c'})
\hspace{5ex} \mbox{or} \hspace{5ex}
(a_1',\ x_{c'} a_1 x_{c'},\ \overline{a_1},\ x_{c'},\ \overline{a_1'})\,.\]
Notice that in the former case, the factorization need not contain $x_{c'}$.  If $\overline{a_1}$ is a complete factor in the factorization of $S_{i - 1}'$, then the diverse palindromic factorization of
\[x_{c'}^3 a_1' x_{c'} a_1 x_{c'} \overline{a_1} x_{c'} \overline{a_1'} x_{c'}^5\]
must include either
\[(x_{c'} a_1' x_{c'},\ a_1,\ x_{c'} \overline{a_1} x_{c'},\ \overline{a_1'})
\hspace{5ex} \mbox{or} \hspace{5ex}
(a_1',\ x_{c'},\ a_1,\ x_{c'} \overline{a_1} x_{c'},\ \overline{a_1'})\,.\]
Again, in the former case, the factorization need not contain $x_{c'}$.  Symmetric propositions hold for $a_2$ and $b$.

We set
\[S_i''
= S_i'\,\$^\dagger \#^\dagger\,x_{c'}^{15} \overline{a_1'} x_{c'} c' x_{c'} \overline{b'} x_{c'}^{17}\,
\$^{\dagger \dagger} \#^{\dagger \dagger}\,x_{c'}^{19} \overline{a_2'} x_{c'} d x_{c'} b' x_{c'}^{21}\,,\]
where $\$^\dagger$, $\#^\dagger$, $\$^{\dagger \dagger}$, $\#^{\dagger \dagger}$, $c'$ and $d$ are symbols we use only here.  Any diverse palindromic factorization of $S_i''$ consists of
\begin{enumerate}
\item a diverse palindromic factorization of $S_i'$,
\item \((\$^\dagger,\ \#^\dagger)\),
\item a diverse palindromic factorization of \(x_{c'}^{15} \overline{a_1'} x_{c'} c' x_{c'} \overline{b'} x_{c'}^{17}\),
\item \((\$^{\dagger \dagger},\ \#^{\dagger \dagger})\),
\item a diverse palindromic factorization of \(x_{c'}^{19} \overline{a_2'} x_{c'} d x_{c'} b' x_{c'}^{21}\).
\end{enumerate}

Since $a_1$ and $a_2$ label outputs in $C_{i - 1}'$ split from the same output in $C_{i - 1}$, it follows that $a_1$ is a complete factor in a diverse palindromic factorization of $S_{i - 1}'$ if and only if $a_2$ is.  Therefore, we need consider only four cases:
%
%
%

\vspace{2mm}

\noindent\textbf{Case 1:} The factorization of $S_{i - 1}'$ includes $a_1$, $a_2$ and $b$ as complete factors, so the factorization of $S_i'$ includes as complete factors either \(x_{c'} \overline{a_1'} x_{c'}\), or $\overline{a_1'}$ and $x_{c'}$; either \(x_{c'} \overline{a_2'} x_{c'}\), or $\overline{a_2'}$ and $x_{c'}$; either \(x_{c'} \overline{b'} x_{c'}\), or $\overline{b'}$ and $x_{c'}$; and $b'$.  Trying all the combinations --- there are only four, since $x_{c'}$ can appear as a complete factor at most once --- shows that any diverse palindromic factorization of $S_i''$ includes one of
\begin{eqnarray*}
&& (\overline{a_1'},\ x_{c'} c' x_{c'},\ \overline{b'},\ \ldots,\ \overline{a_2'},\ x_{c'},\ d,\ x_{c'} b' x_{c'})\,,\\
&& (\overline{a_1'},\ x_{c'} c' x_{c'},\ \overline{b'},\ \ldots,\   \ x_{c'} \overline{a_2'} x_{c'},\ d,\ x_{c'} b' x_{c'})\,,
\end{eqnarray*}
with the latter only possible if $x_{c'}$ appears earlier in the factorization. 
%
%
%


\vspace{2mm}

\noindent\textbf{Case 2:} The factorization of $S_{i - 1}'$ includes $a_1$, $a_2$ and $\overline{b}$ as complete factors, so the factorization of $S_i'$ includes as complete factors either \(x_{c'} \overline{a_1'} x_{c'}\), or $\overline{a_1'}$ and $x_{c'}$; either \(x_{c'} \overline{a_2'} x_{c'}\), or $\overline{a_2'}$ and $x_{c'}$; $\overline{b'}$; and either \(x_{c'} b' x_{c'}\), or $b'$ and $x_{c'}$.  Trying all the combinations shows that any diverse palindromic factorization of $S_i''$ includes one of
\begin{eqnarray*}
&& (\overline{a_1'},\ x_{c'},\ c',\ x_{c'} \overline{b'} x_{c'},\ \ldots,\ \overline{a_2'},\ x_{c'} d x_{c'},\ b')\,,\\
&& (x_{c'} \overline{a_1'} x_{c'},\ c',\ x_{c'} \overline{b'} x_{c'},\ \ldots,\ \overline{a_2'},\ x_{c'} d x_{c'},\ b')\,,
\end{eqnarray*}
with the latter only possible if $x_{c'}$ appears earlier in the factorization. 
%
%
%

\vspace{2mm}

\noindent\textbf{Case 3:} The factorization of $S_{i - 1}'$ includes $\overline{a_1}$, $\overline{a_2}$ and $b$ as complete factors, so the factorization of $S_i'$ includes as complete factors $\overline{a_1'}$; $\overline{a_2'}$; either \(x_{c'} \overline{b'} x_{c'}\), or $\overline{b'}$ and $x_{c'}$; and $b'$.  Trying all the combinations shows that any diverse palindromic factorization of $S_i''$ includes one of
\begin{eqnarray*}
&& (x_{c'} \overline{a_1'} x_{c'},\ c',\ x_{c'},\ \overline{b'},\ \ldots,\ x_{c'} \overline{a_2'} x_{c'},\ d,\ x_{c'} b' x_{c'})\,,\\
&& (x_{c'} \overline{a_1'} x_{c'},\ c',\ x_{c'} \overline{b'} x_{c'},\ \ldots,\ x_{c'} \overline{a_2'} x_{c'},\ d,\ x_{c'} b' x_{c'})\,,
\end{eqnarray*}
with the latter only possible if $x_{c'}$ appears earlier in the factorization.
%
%
%

\vspace{2mm}

\noindent\textbf{Case 4:} The factorization of $S_{i - 1}'$ includes $\overline{a_1}$, $\overline{a_2}$ and $\overline{b}$ as complete factors, so the factorization of $S_i'$ includes as complete factors $\overline{a_1'}$; $\overline{a_2'}$; $\overline{b'}$; and either \(x_{c'} b' x_{c'}\), or $b'$ and $x_{c'}$.   Trying all the combinations shows that any diverse palindromic factorization of $S_i''$ that extends the factorization of $S_i'$ includes one of
\begin{eqnarray*}
&& (x_{c'} \overline{a_1'} x_{c'},\ c',\ x_{c'} \overline{b'} x_{c'},\ \ldots,\ x_{c'} \overline{a_2'} x_{c'},\ d,\ x_{c'},\ b')\,,\\
&& (x_{c'} \overline{a_1'} x_{c'},\ c',\ x_{c'} \overline{b'} x_{c'},\ \ldots,\ x_{c'} \overline{a_2'} x_{c'},\ d,\ x_{c'} b' x_{c'})\,,
\end{eqnarray*}
with the latter only possible if $x_{c'}$ appears earlier in the factorization.
%
%
%

Summing up,  any diverse palindromic factorization of $S_i''$ always includes $x_{c'}$ and includes either \(x_{c'} c' x_{c'}\) if the factorization of $S_{i - 1}'$ includes $a_1$, $a_2$ and $b$ as complete factors, or $c'$ otherwise.

We set
\[S_i''' = S_i''\,\$^{\dagger \dagger \dagger} \#^{\dagger \dagger \dagger}\,x_{c'}^{23} c'' x_{c'} c' x_{c'} \overline{c'} x_{c'} \overline{c''} x_{c'}^{25}\,,\]
where $\$^{\dagger \dagger \dagger}$ and $\#^{\dagger \dagger \dagger}$ are symbols we use only here.  Any diverse palindromic factorization of $S_i'''$ consists of
\begin{enumerate}
\item a diverse palindromic factorization of $S_i''$,
\item \((\$^{\dagger \dagger \dagger},\ \#^{\dagger \dagger \dagger})\),
\item a diverse palindromic factorization of \(x_{c'}^{23} c'' x_{c'} c' x_{c'} \overline{c'} x_{c'} \overline{c''} x_{c'}^{25}\).
\end{enumerate}

Since $x_{c'}$ must appear as a complete factor in the factorization of $S_i''$, if $c'$ is a complete factor in the factorization of $S_i''$, then the factorization of
\[x_{c'}^{23} c'' x_{c'} c' x_{c'} \overline{c'} x_{c'} \overline{c''} x_{c'}^{25}\]
must include
\[(c'',\ x_{c'} c' x_{c'},\ \overline{c'},\ x_{c'} \overline{c''} x_{c'})\,;\]
otherwise, it must include
\[(x_{c'} c'' x_{c'},\ c',\ x_{c'} \overline{c'} x_{c'},\ \overline{c''})\,.\]
That is, the factorization of \(x_{c'}^{23} c'' x_{c'} c' x_{c'} \overline{c'} x_{c'} \overline{c''} x_{c'}^{25}\) includes $c''$, $x_{c'}$ and \(x_{c'} \overline{c''} x_{c'}\) but not $\overline{c''}$ or \(x_{c'} c'' x_{c'}\), if and only if the factorization of $S_i''$ includes $c'$; otherwise, it includes $\overline{c''}$, $x_{c'}$ and \(x_{c'} c'' x_{c'}\) but not $c''$ or \(x_{c'} \overline{c''} x_{c'}\).

We set
\[
S_i = S_i'''\,\$^\ddagger\#^\ddagger\,
x_ccx_cc''x_c\overline{c''}x_c\overline{c}x_c\,,
\] 
where $\$^\ddagger$, $\#^\ddagger$, $c$, $\overline{c}$ and $x_c$ are
symbols that do not appear in $S_i'''$. Any diverse palindromic
factorization of $S_i$ consists of
\begin{enumerate}
\item a diverse palindromic factorization of $S_i'''$,
\item \((\$^{\ddagger},\ \#^{\ddagger})\),
\item a diverse palindromic factorization of 
$x_c c x_cc''x_c\overline{c''}x_c\overline{c}x_c$.
\end{enumerate}

Since exactly one of $c''$ and
$\overline{c''}$ must appear as a complete factor in the factorization
of $S_i'''$, the factorization of
\[
x_ccx_cc''x_c\overline{c''}x_c\overline{c}x_c
\]
must be either
\[
(x_c,\ c,\ x_cc''x_c,\ \overline{c''},\ x_c\overline{c}x_c)
\]
or
\[
(x_ccx_c,\ c'',\ x_c\overline{c''}x_c,\ \overline{c},\ x_c\,).
\]
Thus if $c''$ is a complete factor in the factorization of $S_i'''$, then
$c$, $x_c$ and \(x_c \bar{c} x_c\) are complete factors in the
factorization of $S_i$ but $\bar{c}$, \(x_c c x_c\) and $x_c^j$ are not
for \(j > 1\); otherwise, $\bar{c}$, $x_c$ and \(x_c c x_c\) are
complete factors but $c$, \(x_c \bar{c} x_c\) and $x_c^j$ are not for
\(j > 1\).  

Assume $S_{i - 1}$ represents $C_{i - 1}$.  Let $\tau$ be an
assignment to the inputs of $C_{i - 1}$ and let $P$ be a diverse
palindromic factorization of $S_{i - 1}$ encoding $\tau$.  By
Lemma~\ref{lem:splitting} we can extend $P$ to $P'$ so that it encodes
the assignment to the inputs of $C_{i - 1}'$ that makes them true or
false according to $\tau$.  There are four cases to consider:

\bigskip
\noindent\textbf{Case 1:} $\tau$ makes the outputs of $C_{i -
  1}$ labelled $a$ and $b$ both true. 
Then $P'$ concatenated with, e.g.,
\begin{eqnarray*}
  && (\$,\ \#,\ x_{c'}^3,\ a_1',\ x_{c'} a_1 x_{c'},\ \overline{a_1},\ 
  x_{c'} \overline{a_1'} x_{c'},\ x_{c'}^4,\\ 
  && \$',\ \#',\ x_{c'}^7,\ a_2',\ x_{c'} a_2 x_{c'},\ \overline{a_2},\ 
  x_{c'} \overline{a_2'} x_{c'},\ x_{c'}^8,\\ 
  && \$'',\ \#'',\ x_{c'}^{11},\ b',\ x_{c'} b x_{c'},\ \bar{b},\ 
  x_{c'} \overline{b'} x_{c'},\ x_{c'}^{12})
\end{eqnarray*}
is a diverse palindromic factorization $P''$ of $S_i'$ which,
concatenated with, e.g.,
\begin{eqnarray*}
  && (\$^\dagger,\ \#^\dagger,\ x_{c'}^{15},\ \overline{a_1'},\
  x_{c'} c' x_{c'},\ \overline{b'},\ x_{c'}^{17},\\ 
  && \$^{\dagger \dagger},\ \#^{\dagger \dagger},\ x_{c'}^{19},\ 
  \overline{a_2'},\ x_{c'},\ d,\ x_{c'} b' x_{c'},\ x_{c'}^{20}) 
\end{eqnarray*}
is a diverse palindromic factorization $P'''$ of $S_i''$ which,
concatenated with, e.g.,
\begin{eqnarray*}
  (\$^{\dagger \dagger \dagger},\ \#^{\dagger \dagger \dagger},\
  x_{c'}^{22},\ x_{c'} c'' x_{c'},\ c',\ x_{c'} \overline{c'}
  x_{c'},\ \overline{c''},\ x_{c'}^{25}) 
\end{eqnarray*}
is a diverse palindromic factorization $P^\dagger$ of $S_i'''$ which,
concatenated with
\begin{eqnarray*}
  (\$^{\ddag},\ \#^{\ddag},\ x_{c} c x_{c},\ c'',\ x_{c}
  \overline{c''} x_{c},\ \bar{c},\ x_{c})
\end{eqnarray*}
is  a diverse palindromic factorization $P^\ddagger$ of $S_i$
in which $\bar{c}$, $x_c$ and \(x_c c x_c\) are complete factors but
$c$, \(x_c \bar{c} x_c\) and $x_c^j$ are not for \(j > 1\).

\bigskip
\noindent\textbf{Case 2:} $\tau$ makes the output of $C_{i -
  1}$ labelled $a$ true but the output labelled $b$ false.  Then $P'$
concatenated with, e.g.,
\begin{eqnarray*}
  && (\$,\ \#,\ x_{c'}^3,\ a_1',\ x_{c'} a_1 x_{c'},\ \overline{a_1},\
  x_{c'} \overline{a_1'} x_{c'},\ x_{c'}^4,\\ 
  && \$',\ \#',\ x_{c'}^7,\ a_2',\ x_{c'} a_2 x_{c'},\ \overline{a_2},\
  x_{c'} \overline{a_2'} x_{c'},\ x_{c'}^8,\\ 
  && \$'',\ \#'',\ x_{c'}^{10},\ x_{c'} b' x_{c'},\ b,\ x_{c'} \bar{b}
  x_{c'},\ \overline{b'},\ x_{c'}^{13}) 
\end{eqnarray*}
is a diverse palindromic factorization $P''$ of $S_i'$ which,
concatenated with, e.g.,
\begin{eqnarray*}
  && (\$^\dagger,\ \#^\dagger,\ x_{c'}^{15},\ \overline{a_1'},\
  x_{c'},\ c',\ x_{c'} \overline{b'} x_{c'},\ x_{c'}^{16},\\ 
  && \$^{\dagger \dagger},\ \#^{\dagger \dagger},\ x_{c'}^{19},\ \overline{a_2'},\
  x_{c'} d x_{c'},\ b',\ x_{c'}^{21}) 
\end{eqnarray*}
is a diverse palindromic factorization $P'''$ of $S_i''$ which,
concatenated with, e.g.,
\begin{eqnarray*}
  (\$^{\dagger \dagger \dagger},\ \#^{\dagger \dagger \dagger},\
  x_{c'}^{23},\ c'',\ x_{c'} c' x_{c'},\ \overline{c'},\
  x_{c'} \overline{c''} x_{c'},\ x_{c'}^{24}) 
\end{eqnarray*}
is a diverse palindromic factorization $P^\dagger$ of $S_i'''$ which,
concatenated with
\begin{eqnarray*}
  (\$^{\ddag},\ \#^{\ddag},\ x_{c},\ c,\ x_{c}c''x_{c},\
  \overline{c''},\ x_{c}\bar{c}x_{c})
\end{eqnarray*}
is  a diverse palindromic factorization $P^\ddagger$ of $S_i$
in which $c$, \(x_c \bar{c} x_c\) and $x_c$ are complete factors but
$\bar{c}$, \(x_c c x_c\) and $x_c^j$ are not for \(j > 1\).

\bigskip
\noindent\textbf{Case 3:} $\tau$ makes the output of $C_{i -
  1}$ labelled $a$ false but the output labelled $b$ true.  Then $P'$
concatenated with, e.g.,
\begin{eqnarray*}
  && (\$,\ \#,\ x_{c'}^2,\ x_{c'} a_1' x_{c'},\ a_1,\ x_{c'} \overline{a_1}
  x_{c'},\ \overline{a_1'},\ x_{c'}^5,\\ 
  && \$',\ \#',\ x_{c'}^6,\ x_{c'} a_2' x_{c'},\ a_2,\ x_{c'} \overline{a_2}
  x_{c'},\ \overline{a_2'},\ x_{c'}^9,\\ 
  && \$'',\ \#'',\ x_{c'}^{11},\ b',\ x_{c'} b x_{c'},\ \bar{b},\ 
  x_{c'} \overline{b'} x_{c'},\ x_{c'}^{12})
\end{eqnarray*}
is a diverse palindromic factorization $P''$ of $S_i'$ which,
concatenated with, e.g.,
\begin{eqnarray*}
  && (\$^\dagger,\ \#^\dagger,\ x_{c'}^{14},\ x_{c'} \overline{a_1'}
  x_{c'},\ c',\ x_{c'},\ \overline{b'},\ x_{c'}^{17},\\ 
  && \$^{\dagger \dagger},\ \#^{\dagger \dagger},\ x_{c'}^{18},\
  x_{c'} \overline{a_2'} x_{c'},\ d,\ x_{c'} b' x_{c'},\ x_{c'}^{20}) 
\end{eqnarray*}
is a diverse palindromic factorization $P'''$ of $S_i''$ which,
concatenated with, e.g.,
\begin{eqnarray*}
  (\$^{\dagger \dagger \dagger},\ \#^{\dagger \dagger \dagger},\
  x_{c'}^{23},\ c'',\ x_{c'} c' x_{c'},\ \overline{c'},\
  x_{c'} \overline{c''} x_{c'},\ x_{c'}^{24}) 
\end{eqnarray*}
is a diverse palindromic factorization $P^\dagger$ of $S_i'''$ which,
concatenated with
\begin{eqnarray*}
  (\$^{\ddag},\ \#^{\ddag},\ x_{c},\ c,\ x_{c}c''x_{c},\
  \overline{c''},\ x_{c}\bar{c}x_{c})
\end{eqnarray*}
is  a diverse palindromic factorization $P^\ddagger$ of $S_i$
in which $c$, \(x_c \bar{c} x_c\) and $x_c$ are complete factors but
$\bar{c}$, \(x_c c x_c\) and $x_c^j$ are not for \(j > 1\).

\bigskip
\noindent\textbf{Case 4:} $\tau$ makes the outputs of $C_{i -
  1}$ labelled $a$ and $b$ both false.  Then $P'$
concatenated with, e.g.,
\begin{eqnarray*}
  && (\$,\ \#,\ x_{c'}^2,\ x_{c'} a_1' x_{c'},\ a_1,\ x_{c'} \overline{a_1}
  x_{c'},\ \overline{a_1'},\ x_{c'}^5,\\ 
  && \$',\ \#',\ x_{c'}^6,\ x_{c'} a_2' x_{c'},\ a_2,\ x_{c'} \overline{a_2}
  x_{c'},\ \overline{a_2'},\ x_{c'}^9,\\ 
  && \$'',\ \#'',\ x_{c'}^{10},\ x_{c'} b' x_{c'},\ b,\ x_{c'} \bar{b}
  x_{c'},\ \overline{b'},\ x_{c'}^{13}) 
\end{eqnarray*}
is a diverse palindromic factorization $P''$ of $S_i'$ which,
concatenated with, e.g.,
\begin{eqnarray*}
  && (\$^\dagger,\ \#^\dagger,\ x_{c'}^{14},\ x_{c'} \overline{a_1'}
  x_{c'},\ c',\ x_{c'} \overline{b'} x_{c'},\ x_{c'}^{16},\\ 
  && \$^{\dagger \dagger},\ \#^{\dagger \dagger},\ x_{c'}^{18},\
  x_{c'} \overline{a_2'} x_{c'},\ d,\ x_{c'},\ b',\ x_{c'}^{21}) 
\end{eqnarray*}
is a diverse palindromic factorization $P'''$ of $S_i''$ which,
concatenated with, e.g.,
\begin{eqnarray*}
  (\$^{\dagger \dagger \dagger},\ \#^{\dagger \dagger \dagger},\
  x_{c'}^{23},\ c'',\ x_{c'} c' x_{c'},\ \overline{c'},\
  x_{c'} \overline{c''} x_{c'},\ x_{c'}^{24}) 
\end{eqnarray*}
is a diverse palindromic factorization $P^\dagger$ of $S_i'''$ which,
concatenated with
\begin{eqnarray*}
  (\$^{\ddag},\ \#^{\ddag},\ x_{c},\ c,\ x_{c}c''x_{c},\
  \overline{c''},\ x_{c}\bar{c}x_{c})
\end{eqnarray*}
is  a diverse palindromic factorization $P^\ddagger$ of $S_i$
in which $c$, \(x_c \bar{c} x_c\) and $x_c$ are complete factors but
$\bar{c}$, \(x_c c x_c\) and $x_c^j$ are not for \(j > 1\).

\bigskip

Notice that in all cases $P^\ddagger$ encodes the assignment to the
inputs of $C_i$ that makes them true or false according to $\tau$.
Since $C_{i - 1}$ and $C_i$ have the same inputs, each assignment to
the inputs of $C_i$ is encoded by some diverse palindromic
factorization of $S_i$.

Now let $P$ be a diverse palindromic factorization of $S_i$ and let $\tau$ be the assignment to the inputs of $C_{i - 1}$ that is encoded by a prefix of $P$. 
Let $\hat{P}$ be a diverse palindromic factorization of $S_{i-1}'$. Since $a_1$ and $a_2$ are obtained by splitting $a$ in $S_{i-1}$, it follows that $a_1$ is a complete factor of $\hat{P}$ if and only if $a_2$ is. Therefore, in what follows we only consider any diverse palindromic factorization $P$ of $S_{i}$ in which either both $a_1$ and $a_2$ are complete factors, or neither $a_1$ nor $a_2$ is a complete factor.

Let $P'$ be the prefix of $P$ that is a diverse palindromic factorization of $S_i'''$.

\noindent\textbf{Case A:}
Suppose the factorization of
\[x_{c'}^{23} c'' x_{c'} c' x_{c'} \overline{c'} x_{c'} \overline{c''} x_{c'}^{25}\]
in $P'$ includes $\overline{c''}$ as a complete factor, which is the case if and only if $P$ includes $\bar{c}$, $x_c$ and \(x_c c x_c\) as complete factors but not $c$, \(x_c \bar{c} x_c\) and $x_c^j$ for \(j > 1\).  We will show that $\tau$ must make the outputs of $C_{i - 1}$ labelled $a$ and $b$ true. 
Let $P''$ be the prefix of $P'$ that is a diverse palindromic factorization of $S_i''$.  Since $\overline{c''}$ is a complete factor in the factorization of
\[x_{c'}^{23} c'' x_{c'} c' x_{c'} \overline{c'} x_{c'} \overline{c''} x_{c'}^{25}\]
in $P'$, so is $c'$.  Therefore, $c'$ is not a complete factor in the factorization of
\[x_{c'}^{15} \overline{a_1'} x_{c'} c' x_{c'} \overline{b'} x_{c'}^{17}\]
in $P''$, so $\overline{a_1'}$ and $\overline{b'}$ are.

Let $P'''$ be the prefix of $P''$ that is a diverse palindromic factorization of $S_i'$.  Since $\overline{a_1'}$ and $\overline{b'}$ are complete factors later in $P''$, they are not complete factors in $P'''$.  Therefore, $\overline{a_1}$ and $\bar{b}$ are complete factors in the factorizations of
\[x_{c'}^3 a_1' x_{c'} a_1 x_{c'} \overline{a_1} x_{c'} \overline{a_1'} x_{c'}^5
\hspace{5ex} \mbox{and} \hspace{5ex}
x_{c'}^{11} b' x_{c'} b x_{c'} \bar{b} x_{c'} \overline{b'} x_{c'}^{13}\]
in $P'''$, so they are not complete factors in the prefix $P^\dagger$ of $P$ that is a diverse palindromic factorization of $S_{i - 1}'$.  Since we built $S_{i - 1}'$ from $S_{i - 1}$ with Lemma~\ref{lem:splitting}, it follows that $a_1$ and $b$ are complete factors in the prefix of $P$ that encodes $\tau$.  Therefore, $\tau$ makes the outputs of $C_{i - 1}$ labelled $a$ and $b$ true.

\noindent\textbf{Case B:}
Suppose the factorization of
\[x_{c'}^{23} c'' x_{c'} c' x_{c'} \overline{c'} x_{c'} \overline{c''} x_{c'}^{25}\]
in $P'$ does not include $\overline{c''}$ as a complete factor,
which implies that it does include $x_{c'} \overline{c''} x_{c'}$ as a complete factor.
Since, as noted earlier, we can assume that $a_1$ is a complete factor of $P$ if and only if $a_2$ is,
it follows
that the factorization of
\[x_{c'}^{23} c'' x_{c'} c' x_{c'} \overline{c'} x_{c'} \overline{c''} x_{c'}^{25}\] must include
\[(c'', x_{c'} c' x_{c'}, \overline{c'}, x_{c'} \overline{c''} x_{c'}).\]
Then, $P$ must include $x_c$, $c$ and $\overline{c''}$ as complete factors.
We will show that $\tau$ must make at least one of the outputs of $C_{i - 1}$ labelled $a$ or $b$ false.
Let $P''$ be the prefix of $P'$ that is a diverse palindromic factorization of $S_i''$.
Since $x_{c'} c' x_{c'}$  is a complete factor in the factorization of
\[x_{c'}^{23} c'' x_{c'} c' x_{c'} \overline{c'} x_{c'} \overline{c''} x_{c'}^{25}\]
in $P'$, $c'$ is a complete factor in the factorization of
\[x_{c'}^{15} \overline{a_1'} x_{c'} c' x_{c'} \overline{b'} x_{c'}^{17}\]
in $P''$.
Then, the factorization of 
\[x_{c'}^{15} \overline{a_1'} x_{c'} c' x_{c'} \overline{b'} x_{c'}^{17}\] 
must include one of the following three:
\begin{eqnarray}
  & (x_{c'} \overline{a_1'} x_{c'}, c', x_{c'} \overline{b'} x_{c'}), &  \label{CaseBa} \\
  & (x_{c'} \overline{a_1'} x_{c'}, c', x_{c'}, \overline{b'}),&  \label{CaseBb} \\
  & (\overline{a_1'}, x_{c'}, c', x_{c'} \overline{b'} x_{c'}). & \label{CaseBc}
\end{eqnarray}
\begin{description}
\item[Case B-a:]
  Assume the factorization of $x_{c'}^{15} \overline{a_1'} x_{c'} c' x_{c'} \overline{b'} x_{c'}^{17}$ includes~(\ref{CaseBa}).
Let $P'''$ be the prefix of $P''$ that is a diverse palindromic factorization of $S_i'$.
Since $\overline{a_1'}$ and $\overline{b'}$ are not complete factors later in $P''$, they are complete factors in $P'''$.
Therefore, there are five combinations of factorizations of 
\[x_{c'}^3 a_1' x_{c'} a_1 x_{c'} \overline{a_1} x_{c'} \overline{a_1'} x_{c'}^5
\hspace{5ex} \mbox{and} \hspace{5ex}
x_{c'}^{11} b' x_{c'} b x_{c'} \bar{b} x_{c'} \overline{b'} x_{c'}^{13}\]
in $P'''$, as follows:\par
\begin{description}
\item[Case B-a1:]
The factorizations include 
\[(x_{c'} a_1' x_{c'},\ a_1,\ x_{c'} \overline{a_1} x_{c'},\ \overline{a_1'}) \mbox{ and }
(x_{c'} b' x_{c'},\ b,\ x_{c'} \bar{b} x_{c'},\ \overline{b'}).\]
In this case, $a_1$ and $b$ are not complete factors in the prefix of $P$ that encodes $\tau$. Therefore, $\tau$ makes both the outputs of $C_{i-1}$ labelled $a$ and $b$ false.\par
\item[Case B-a2:]
The factorizations include
\[ (x_{c'} a_1' x_{c'},\ a_1,\ x_{c'} \overline{a_1} x_{c'},\ \overline{a_1'}) \mbox{ and }
(b',\ x_{c'} b x_{c'},\ \bar{b},\ x_{c'},\ \overline{b'}).\]
In this case, $a_1$ is not a complete factor and $b$ is a complete factor in the prefix of $P$ that encodes $\tau$. Therefore, $\tau$ makes the outputs of $C_{i-1}$ labelled $a$ false and $b$ true.\par
\item[Case B-a3:]
The factorizations include 
\[ (a_1',\ x_{c'} a_1 x_{c'},\ \overline{a_1},\ x_{c'},\ \overline{a_1'})  \mbox{ and }
 (x_{c'} b' x_{c'},\ b,\ x_{c'} \bar{b} x_{c'},\ \overline{b'}).\]
In this case, $a_1$ is a complete factor and $b$ is not a complete factor in the prefix of $P$ that encodes $\tau$. Therefore, $\tau$ makes the outputs of $C_{i-1}$ labelled $a$ true and $b$ false.\par
\item[Case B-a4:]
The factorizations include \[ (a_1',\ x_{c'},\ a_1,\ x_{c'} \overline{a_1} x_{c'},\ \overline{a_1'}) \mbox{ and }
(x_{c'} b' x_{c'},\ b,\ x_{c'} \bar{b} x_{c'},\ \overline{b'}).\]
In this case, $a_1$ and $b$ are not complete factors in the prefix of $P$ that encodes $\tau$. Therefore, $\tau$ makes both the outputs of $C_{i-1}$ labelled $a$ and $b$ false.\par
\item[Case B-a5:]
The factorizations include \[(x_{c'} a_1' x_{c'},\ a_1,\ x_{c'} \overline{a_1} x_{c'},\ \overline{a_1'}) \mbox{ and }
(b',\ x_{c'},\ b,\ x_{c'} \bar{b} x_{c'},\ \overline{b'}).\]
In this case, $a_1$ and $b$ are not complete factors in the prefix of $P$ that encodes $\tau$. Therefore, $\tau$ makes both the outputs of $C_{i-1}$ labelled $a$ and $b$ false.
\end{description}
\item[Case B-b:]
  Assume the factorization of $x_{c'}^{15} \overline{a_1'} x_{c'} c' x_{c'} \overline{b'} x_{c'}^{17}$ includes~(\ref{CaseBb}).
Let $P''$ be the prefix of $P'$ that is a diverse palindromic factorization of $S_i''$.
Let $P'''$ be the prefix of $P''$ that is a diverse palindromic factorization of $S_i'$.
Since $\overline{a_1'}$ and $x_{c'} \overline{b'} x_{c'}$ are not complete factors later in $P''$, they are complete factors in $P'''$.
Therefore, the factorizations of
\[x_{c'}^3 a_1' x_{c'} a_1 x_{c'} \overline{a_1} x_{c'} \overline{a_1'} x_{c'}^5
\hspace{5ex} \mbox{and} \hspace{5ex}
x_{c'}^{11} b' x_{c'} b x_{c'} \bar{b} x_{c'} \overline{b'} x_{c'}^{13}\]
must include
\[(x_{c'} a_1' x_{c'},\ a_1,\ x_{c'} \overline{a_1} x_{c'},\ \overline{a_1'})
\mbox{ and }
(b',\ x_{c'} b x_{c'},\ \bar{b},\ x_{c'} \overline{b'} x_{c'})\]
in $P'''$.
Then $a_1$ is not a complete factor and $b$ is a complete factor in the prefix of $P$ that encodes $\tau$. Therefore, $\tau$ makes the outputs of $C_{i-1}$ labelled $a$ false and $b$ true.
\item[Case B-c:]
Assume the factorization of $x_{c'}^{15} \overline{a_1'} x_{c'} c' x_{c'} \overline{b'} x_{c'}^{17}$ includes~(\ref{CaseBc}).
Let $P''$ be the prefix of $P'$ that is a diverse palindromic factorization of $S_i''$.
Let $P'''$ be the prefix of $P''$ that is a diverse palindromic factorization of $S_i'$.
Since $x_{c'} \overline{a_1'} x_{c'}$ and $\overline{b'}$ are not complete factors later in $P''$, they are complete factors in $P'''$.
Therefore, the factorizations of
\[x_{c'}^3 a_1' x_{c'} a_1 x_{c'} \overline{a_1} x_{c'} \overline{a_1'} x_{c'}^5
\hspace{5ex} \mbox{and} \hspace{5ex}
x_{c'}^{11} b' x_{c'} b x_{c'} \bar{b} x_{c'} \overline{b'} x_{c'}^{13}\]
must include
\[(a_1',\ x_{c'} a_1 x_{c'},\ \overline{a_1},\ x_{c'} \overline{a_1'} x_{c'})
\mbox{ and }
(x_{c'} b' x_{c'},\ b,\ x_{c'} \bar{b} x_{c'},\ \overline{b'})\]
in $P'''$.
Then $a_1$ is a complete factor and $b$ is not a complete factor in the prefix of $P$ that encodes $\tau$. Therefore, $\tau$ makes the outputs of $C_{i-1}$ labelled $a$ true and $b$ false.
\end{description}

The above arguments give the following lemma.

\begin{lemma}
\label{lem:nanding}
If we have a string $S_{i - 1}$ that represents $C_{i - 1}$ and $C_i$ is obtained from $C_{i - 1}$ by making two outputs of $C_{i - 1}$ the inputs of a new NAND gate, then in constant time we can append symbols to $S_{i - 1}$ to obtain a string $S_i$ that represents~$C_i$.
\end{lemma}


\section{Summing Up}
\label{sec:conclusion}

By Lemmas~\ref{lem:adding},~\ref{lem:splitting} and~\ref{lem:nanding} and induction, given a Boolean circuit $C$ composed only of splitters and NAND gates with two inputs and one output, in time linear in the size of $C$ we can build, inductively and in turn, a sequence of strings \(S_1, \ldots, S_t\) such that $S_i$ represents $C_i$.  As mentioned in Section~\ref{sec:outline}, once we have $S_t$ we can easily build in constant time a string $S$ that has a diverse palindromic factorization if and only if $C$ is satisfiable.  Therefore, diverse palindromic factorization is NP-hard.  Since it is obviously in NP, we have the following theorem:

\begin{theorem}
\label{thm:conclusion}
Diverse palindromic factorization is NP-complete.
\end{theorem}

\section{$k$-Diverse Factorization}
\label{sec:kdiverse}

It is not difficult to check that our reduction is still correct even if factors of the forms $\$$, $\#$ and $x^j$ for \(j > 1\) can appear arbitrarily often in the factorization, as long as factors of the forms $a$, $x$ and \(x a x\) can each appear at most once.  (By ``of the form'' we mean equal up  to subscripts, bars and superscripts apart from exponents; $a$ stands for any letter except $x$.)  It follows that it is still NP-complete to decide for any fixed $k$ whether a string can be factored into palindromes that each appear at most $k$ times in the factorization.

Suppose we are given $k$ and a Boolean circuit $C$ composed only of splitters and NAND gates with two inputs and one output.  In linear time we can build, as we have described, a string $S$ such that $S$ has a diverse palindromic factorization if and only if $C$ is satisfiable.  In linear time we can then build a string $T$ as follows: we start with $T$ equal to the empty string; for each substring of $S$ of the form $a$, we append to $T$ a substring of the form
\[\$_1 \#_1\,a\,\$_2 \#_2\,a\,\$_3 \#_3 \cdots \$_{k - 1} \#_{k - 1}\,a\,\$_k \#_k\,,\]
where \(\$_1, \ldots, \$_k, \#_1, \ldots, \#_k\) are symbols we use only here; for each substring of $S$ of the form $x$, we append to $T$ a substring of the form
\[\$_1' \#_1'\,x\,\$_2' \#_2'\,x\,\$_3' \#_3' \cdots \$_{k - 1}' \#_{k - 1}'\,x\,\$_k' \#_k'\,,\]
where \(\$_1', \ldots, \$_k', \#_1', \ldots, \#_k'\) are symbols we use only here; for each substring of $S$ of the form \(x a x\), we append to $T$ a substring of the form
\[\$_1'' \#_1''\,x a x\,\$_2'' \#_2''\,x a x\,\$_3'' \#_3'' \cdots \$_{k - 1}'' \#_{k - 1}''\,x a x\,\$_k'' \#_k''\,,\]
where \(\$_1'', \ldots, \$_k'', \#_1'', \ldots, \#_k''\) are symbols we use only here.

Notice that the only $k$-diverse palindromic factorization of $T$ 
includes each substring of $S$ of the
forms $a$, $x$ and \(x a x\) exactly \(k - 1\) times each.  In
particular, any substring of $T$ of the form \(x a x\) cannot be
factored into \((x,\ a,\ x)\), because $x$ must appear \(k - 1\) times
elsewhere in the factorization.  Therefore, there is a $k$-diverse
palindromic factorization of \(S\,\$ \#\,T\),
where $\$$ and $\#$ are symbols we use only here, if and
only if there is a diverse palindromic factorization of $S$ and, thus,
if and only if $C$ is satisfiable.  This implies the following
generalization of Theorem~\ref{thm:conclusion}.

\begin{theorem}
\label{thm:kdiverse}
For any fixed $k\ge 1$, $k$-diverse palindromic factorization is NP-complete.
\end{theorem}


\section{Binary Alphabet}
\label{sec:binary}

The reduction described above involves multiple distinct symbols for
each component of the circuit and thus requires an unbounded
alphabet, but we will next show that a binary alphabet is sufficient.

Let $S$ be an arbitrary string and let $\Sigma$ be the set of distinct
symbols occurring in $S$. Let $\delta$ be an (arbitrary) bijective
mapping $\delta : \Sigma \to \{ba^ib : i\in[1..|\Sigma|]\}$.  We
will also use $\delta$ to denote the implied mapping from $\Sigma^*$
to $\{a,b\}^*$ defined recursively by $\delta(X\alpha) = \delta(X)
\cdot \delta(\alpha)$ for any $X\in\Sigma^*$ and $\alpha\in\Sigma$.

Notice that $\delta$ preserves palindromes, i.e., for any palindrome
$P\in\Sigma^*$, $\delta(P)$ is a palindrome too.  Thus, if
$\mathbf{P}=(P_1, P_2, \dots, P_k)$ is a palindromic factorization of
$S$, then $\delta(\mathbf{P})=(\delta(P_1), \delta(P_2), \dots,
\delta(P_k))$ is a palindromic factorization of $\delta(S)$.
Furthermore any palindrome in $\delta(S)$ of the form $(ba^+b)^+$ must
be a preserved palindrome, i.e., an image $\delta(P)$ of a palindrome
$P$ occurring in $S$. Any palindromic factorization of $\delta(S)$
consisting of preserved palindromes only corresponds to a palindromic
factorization of $S$. We call this a preserved palindromic
factorization of $\delta(S)$. Notice that a preserved palindromic
factorization $\delta(\mathbf{P})$ is diverse if and only if
$\mathbf{P}$ is diverse.

Now consider an arbitrary non-preserved palindromic factorization of
$\delta(S)$.  It is easy to see that the first palindrome must be
either a single $b$ or a preserved palindrome. Furthermore, any
palindrome following a preserved palindrome in the factorization must
be either a single $b$ or a preserved palindrome. Thus the palindromic
factorization of $\delta(S)$ begins with a (possibly empty) sequence
of preserved palindromes followed by a single $b$.  A symmetric
argument shows that the factorization also ends with a (possibly
empty) sequence of preserved palindromes preceded by a single $b$.
The two single $b$'s cannot be the same $b$ since one is the first $b$
in an image of a symbol in $S$, and the other is a last $b$. 
Thus a non-preserved palindromic factorization can never be diverse.

The above discussion proves the following lemma.

\begin{lemma}
  For any string $S$, $\delta(S)$ has a diverse palindromic
  factorization if and only if $S$ has a diverse palindromic factorization.
\end{lemma}

Applying the lemma to the string $S$ constructed from a Boolean
circuit $C$ as described in Sections~\ref{sec:adding},
\ref{sec:splitting} and \ref{sec:nanding}, shows that $\delta(S)$ has
a diverse palindromic factorization if and only if $C$ is satisfiable.
Since $\delta(S)$ can be constructed in time quadratic in the size of
$C$, we have a binary alphabet version of
Theorem~\ref{thm:conclusion}.

\begin{theorem}
\label{thm:binary}
  Diverse palindromic factorization of binary strings is NP-complete.
\end{theorem}

If we allow each factor to occur at most $k>1$ times, the above
transformation to a binary alphabet does not work anymore, because two
single $b$'s is now allowed. However, a small modification is
sufficient to correct this.  First, we replace $\delta$ with a
bijection $\delta' : \Sigma \to \{ba^ib :
i\in[3..|\Sigma|+2]\}$. Second, we append to $\delta'(S)$ the string
$Q_k$ which is a length $20k$ prefix of $(abbaab)^*$.

Let us first analyze the palindromic structure of $Q_k$.
It is easy to see that the only palindromes in $Q_k$ are 
\[
a,\ b,\ aa,\ bb,\ aba,\ bab,\ abba,\ \textrm{and}\ baab.
\]
The total length of these palindromes is 20 and thus the only possible
$k$-diverse palindromic factorization of $Q_k$
is one where all the above palindromes appear exactly $k$
times. Such factorizations exist too. For example, $k$ copies of
\[
(abba,\ aba,\ bb,\ aa,\ bab,\ baab)
\]
followed by $2k$ single symbol palindromes is such a factorization.

Now consider the string $\delta'(S)Q_k$. It is easy to verify that the
only palindromes overlapping both $\delta'(S)$ and $Q_k$ are $aba$ and
$bab$. However, in any palindromic factorization containing one of
them, the factorization of the remaining part of $Q_k$ together with
the overlapping palindrome would have to contain more than $k$
occurrences of some factor.  Thus in any $k$-diverse palindromic
factorization of $\delta'(S)Q_k$, there are no overlapping palindromes
and the factorizations of $\delta'(S)$ and $Q_k$ are separate.  Since
the factorization of $Q_k$ contains $k$ single $b$'s, the
factorization of $\delta'(S)$ cannot contain any single $b$'s. Then,
by the discussion earlier in this section, all palindromes in
$\delta'(S)$ must be preserved palindromes.

\begin{lemma}
  For any string $S$ and any $k\ge 1$, the string $\delta'(S)Q_k$ has
  a $k$-diverse palindromic factorization if and only if $S$ has a
  $k$-diverse palindromic factorization.
\end{lemma}

Combining this with Theorem~\ref{thm:kdiverse}, we obtain the
following:

\begin{theorem}
\label{thm:kdiverse-binary}
  For any fixed $k \ge 1$, $k$-diverse palindromic factorization of
  binary strings is NP-complete.
\end{theorem}

\section*{Acknowledgments}

Many thanks to Gabriele Fici for his comments on a draft of this paper, and to the anonymous referee who pointed out a gap in the proof of Lemma~\ref{lem:nanding}.

This work has been supported by grants 268324, 258308 and 284598 from
the Academy of Finland, and by a research fellowship within the
project ``Enhancing Educational Potential of Nicolaus Copernicus
University in the Disciplines of Mathematical and Natural Sciences''
(project no.\ POKL.04.01.01-00-081/10).

This work was done while the second author was at the Department of Computer Science and the Helsinki Institute for Information Technology (HIIT) of the University of Helsinki.

\bibliographystyle{ws-ijfcs}
\bibliography{diverse}

\begin{thebibliography}{10}

\bibitem{AIR13}
A.~Alitabbi, C.~S. Iliopoulos and M.~S. Rahman, Maximal palindromic
  factorization, {\em Proceedings of the Prague Stringology Conference
  (PSC)\/},   (2013), pp. 70--77.

\bibitem{BGIKKPPS15}
H.~Bannai, T.~Gagie, S.~Inenaga, J.~K{\"a}rkk{\"a}inen, D.~Kempa,
  M.~Pi\k{a}tkowski, S.~J. Puglisi and S.~Sugimoto, Diverse palindromic
  factorization is {NP}-complete, {\em Proceedings of the 19th Conference on
  Developments in Language Theory (DLT)\/},   (2015), pp. 85--96.

\bibitem{BS14}
S.~Buss and M.~Soltys, Unshuffling a square is {NP}-hard, {\em Journal of
  Computer and System Sciences} {\bf 80}(4)  (2014)  766--776.

\bibitem{CFGGS16}
K.~Casel, H.~Fernau, S.~Gaspers, B.~Gras and M.~L. Schmid, On the complexity of
  grammar-based compression over fixed alphabets, {\em Proceedings of the 43rd
  International Colloquium on Automata, Languages, and Programming (ICALP)\/},
   (2016), pp. 122:1--122:14.

\bibitem{FMMS15}
H.~Fernau, F.~Manea, R.~Merca\c{s} and M.~L. Schmid, Pattern matching with
  variables: Fast algorithms and new hardness results, {\em Proceedings of the
  32nd Symposium on Theoretical Aspects of Computer Science (STACS)\/},
  (2015), pp. 302--315.

\bibitem{FGKK14}
G.~Fici, T.~Gagie, J.~K{\"a}rkk{\"a}inen and D.~Kempa, A subquadratic algorithm
  for minimum palindromic factorization, {\em Journal of Discrete Algorithms}
  {\bf 28}  (2014)  41--48.

\bibitem{FPZ13}
A.~E. Frid, S.~Puzynina and L.~Zamboni, On palindromic factorization of words,
  {\em Advances in Applied Mathematics} {\bf 50}(5)  (2013)  737--748.

\bibitem{GJ79}
M.~R. Garey and D.~S. Johnson, {\em Computers and Intractability: A Guide to
  the Theory of {NP}-Completeness} (W. H. Freeman and Co., 1979).

\bibitem{GMSU16}
P.~Gawrychowski, O.~Merkurev, A.~M. Shur and P.~Uznanski, Tight tradeoffs for
  real-time approximation of longest palindromes in streams, {\em Proceedings
  of the 27th Symposium on Combinatorial Pattern Matching (CPM)\/},   (2016),
  pp. 18:1--18:13.

\bibitem{HLR16}
D.~Hucke, M.~Lohrey and C.~P. Reh, The smallest grammar problem revisited, {\em
  Proceedings of the 23rd Symposium on String Processing and Information
  Retrieval (SPIRE)\/},   (2016), pp. 35--49.

\bibitem{ISIBT14}
T.~I, S.~Sugimoto, S.~Inenaga, H.~Bannai and M.~Takeda, Computing palindromic
  factorizations and palindromic covers on-line, {\em Proceedings of the 25th
  Symposium on Combinatorial Pattern Matching (CPM)\/},   (2014), pp. 150--161.

\bibitem{KRS15}
D.~Kosolobov, M.~Rubinchik and A.~M. Shur, Pal$^k$ is linear recognizable
  online, {\em Proceedings of the 41st Conference on Current Trends in Theory
  and Practice of Computer Science (SOFSEM)\/},   (2015), pp. 289--301.

\bibitem{Lev73}
L.~Levin, Universal search problems, {\em Problems of Information Transmission}
  {\bf 9}(3)  (1973)  115--116.

\bibitem{Rav03}
O.~Ravsky, On the palindromic decomposition of binary words, {\em Journal of
  Automata, Languages and Combinatorics} {\bf 8}(1)  (2003)  75--83.

\bibitem{Sch16}
M.~L. Schmid, Computing equality-free and repetitive string factorizations,
  {\em Theoretical Computer Science} {\bf 618}(7)  (2016)  42--51.

\bibitem{Tse68}
G.~S. Tseitin, On the complexity of derivation in propositional calculus, {\em
  Structures in Constructive Mathematics and Mathematical Logic, Part II\/},
  ed. A.~O. Slisenko 1968, pp. 115--125.

\bibitem{ZL77}
J.~Ziv and A.~Lempel, A universal algorithm for sequential data compression,
  {\em IEEE Transactions on Information Theory} {\bf 22}(3)  (1977)  337--343.

\bibitem{ZL78}
J.~Ziv and A.~Lempel, Compression of individual sequences via variable-rate
  coding, {\em IEEE Transactions on Information Theory} {\bf 24}(5)  (1978)
  530--536.

\end{thebibliography}

\end{document}